 \documentclass[aps,pre,showpacs,twocolumn,superscriptaddress,floatfix]{revtex4-1}
\usepackage{graphicx}
\usepackage{color}
\usepackage{amsmath,amsfonts,amssymb,wasysym}
\usepackage{physics}

\definecolor{myblue}{rgb}{0.368417, 0.506779, 0.709798}
\definecolor{myyellow}{rgb}{0.880722, 0.611041, 0.142051}
\definecolor{mygreen}{rgb}{0.560181, 0.691569, 0.194885}
\definecolor{myred}{rgb}{0.922526, 0.385626, 0.209179}
\begin{document}
\title{
Mean first passage times reconstruct
the slowest relaxations in potential energy landscapes of nanoclusters
}
\author{Teruaki Okushima}
\email{okushima@isc.chubu.ac.jp}
\affiliation{College of Engineering,
Chubu University, Matsumoto-cho, Kasugai, Aichi 487-8501, Japan}
\author{Tomoaki Niiyama}
\email{niyama@se.kanazawa-u.ac.jp}
\affiliation{
Graduate School of Natural Science and Technology, Kanazawa University,
Kakuma-cho, Kanazawa, Ishikawa 920-1192, Japan
}
\author{Kensuke S. Ikeda}
\email{ahoo@ike-dyn.ritsumei.ac.jp}
\affiliation{
College of Science and Engineering, Ritsumeikan University, 
Noji-higashi 1-1-1, Kusatsu, shiga 525-8577, Japan}
\author{Yasushi Shimizu}
\email{shimizu@se.ritsumei.ac.jp}
\affiliation{Department of Physics, Ritsumeikan University,
Noji-higashi 1-1-1, Kusatsu, shiga 525-8577, Japan}
\date{\today}
\begin{abstract} 
Relaxation modes are the collective modes in which
all probability deviations from equilibrium states 
decay with the same relaxation rates.
In contrast,
a first passage time is 
the required time for arriving for the first time
from one state to another.
 In this paper,
 we discuss how and why the slowest relaxation rates 
 of relaxation modes are reconstructed from the  first passage times.
 As an illustrative model,
 we use 
 a continuous-time  Markov state model of
 vacancy diffusion in KCl nanoclusters.
Using this model,
we reveal that
all characteristics of the relaxations in KCl nanoclusters
come from the fact
that they are hybrids of 
two kinetically different regions of
 the fast surface  and slow bulk diffusions.
The origin of the different diffusivities turns out to come from
the heterogeneity of the activation energies on
 the potential energy landscapes.
We also develop a stationary population method
to compute the mean first passage times as
mean times required for pair annihilations of particle-hole pairs,
which enables us
to obtain 
the symmetric results of relaxation rates under
the exchange of
the sinks and  the sources.
 With this symmetric method,
we finally show 
why the slowest relaxation times can be reconstructed from
the mean first passage times.
\end{abstract}

\maketitle
\section{Introduction}

Recently,
the dynamics of complex systems, 
such as
the relaxation of glass-forming materials
\cite{goldstein,stillingerWeber2,stillinger,heuer,APRV,debenedettiStillinger,
sastory,DRB,DH,AFMK,DSSSKG,yang},
the kinetics of biomolecules \cite{BeckerKerplus,folding,kinesin,gene,hummer,hummer2,wolynes,pande,wang,kern,nucleosome}, and
diffusion in nanoclusters \cite{bbskj,hhdksi,jkk,hshias,hbssh,cubic,niiyama},
were studied in a unified way for Markov state models \cite{wales0,wales,stillingerBook,BPE,lumping}.
The slowest relaxation modes of these systems
describe
the bottleneck processes, and hence
they are the most crucial,
e.g.,
for understanding 
glass transitions and 
rapid formations of mixed crystals
\cite{clumping,clumping2,dps,kfs,maeno,oku2007,rg,rgtime}.

The relaxation rates and modes are
the eigenvalues and eigenvectors, respectively,
of the transition rate matrix of a Markov state model.
In general,
physical quantities are expressed in terms of the eigenvalues and eigenvectors.
The resulting expressions,
called spectral representations,
give useful formulas that enable us to
evaluate the physical quantities 
with use of
the eigenvalues and eigenvectors
\cite{reresub1,reresub2,reresub3,reresub4,reresub5,reresub6,ngvankampenResub,HartichGodecResub,aldousResub}.
We can compute the relaxation rates and modes 
of realistic, complicated Markov state models 
using numerical diagonalization algorithms \cite{matrixcomp}.
However,
it is hard to understand why the eigenvectors are formed in the shapes of
the numerical diagonalization results
because the eigenvectors are quite high-dimensional and complicated.
To extract the essence of the relaxation properties of
Markov state models,
there have been many studies,
such as
lumping or renormalizing Markov state models \cite{lumping,clumping,clumping2,dps,kfs,rg,rgtime},
and
applications of
network algorithms,
such as Dijkstra's shortest path algorithm \cite{maeno}.
Although there are many pioneering works concerning this problem 
\cite{reresub1,reresub2,reresub3,reresub4,reresub5,reresub6,ngvankampenResub,HartichGodecResub,aldousResub},
to the best of our knowledge, this problem has not yet been completely clarified.

As a more specific indicator of diffusive transport
than the slowest relaxation rates, 
the first passage time is widely studied
mainly for analyzing the kinetics in complex networks
\cite{mfpt,ddl,dsl,lhl,tbv,kinesin,gene}.
The first passage times from one state to another target state
in a kinetic network
are 
the required times of stochastic realizations
for traveling from the former to the latter state for the first time. 
The corresponding mean first passage time is
given by 
the mean value of the first passage times of the stochastic realizations.

Intuitively,
we may interpret the slowest relaxation of a system
as the process that transports the excess probability
to the  maximum probability states of the equilibrium distribution
along the unavoidable and slowest transport routes
in order to achieve the equilibrium distribution. 
Therefore, it may be possible to understand the formation of the slowest relaxation mode by searching for the states, where the first passage times to the maximum probability states are the largest, and then by finding out the main routes connecting the former to the latter states.
To the best of our knowledge, however,
there have been no such studies that search for the slowest relaxations
with this idea.
Instead, all pioneering works, e.g., Refs.~\cite{reresub1,reresub2,reresub3,reresub4,reresub5,reresub6,ngvankampenResub,HartichGodecResub}, concern mainly how the mean first passage times are expressed with the relaxation modes via renewal theorems. 
It should be noted that  in this paper
we study the inverse problem,
i.e.,
how and why the slowest relaxation modes are reconstructed by the mean first passage times.

As a realistic problem,
we analyze a KCl nanocluster model
having one vacancy \cite{kimura,cubic,niiyama}.
The vacancy diffuses in the cluster
and introduces
the mixing of atoms in the cluster.
As for the pioneering works  on vacancy diffusion,
the equilibrium vacancy concentration \cite{nt1resub}
and
the relaxation process using a stochastic process simplified by a
 uniform diffusion equation \cite{nt2resub} have been studied.
Nevertheless,
there are no studies 
in which the surface effects of nanoclusters on the relaxations
of the vacancy diffusion
are taken into consideration. 
The most substantial reason that makes such  approaches difficult is
that it is hard
to estimate the transition rates between all adjacent states
on the high-dimensional potential energy surface
from interatomic interactions.
Fortunately,
in Ref.~\cite{niiyama}
we have successfully enumerated all states and
all transition rates between adjacent states
in nanoclusters of various sizes,
and we elucidated the specific properties,
 such as migration energies of vacancies,
 arising from  the surfaces of nanoclusters.
In this paper, we use these transition data
to construct the Markov state model of KCl nanoclusters
and investigate the relationship
between the slowest relaxation modes
and the first passage times in the Markov state model
equipped with the cluster surfaces of 
characteristic transition regions.

The purpose of this paper is twofold.
One is
to understand  the formations of the slowest relaxation modes
in terms of the first passage times of 
the vacancy diffusion in KCl nanoclusters.
The other is
to elucidate the theoretical basis for
why such a mean first passage time analysis
applies to the slowest relaxations.

In Sec.~\ref{sec:mfpts},
we introduce
 a Markov state model,
  its relaxation modes,
  and
  its mean first passage times in a general setting.
For the mean first passage times,
we develop a stationary population method
that enables us to compute the first passage times
from the stationary populations of
the Markov state models that connect sinks with sources.
In Sec.~\ref{sec:model},
we introduce the interatomic interaction of the KCl nanoclusters
and then  construct the Markov state models of the vacancy diffusion model.
In Sec.~\ref{sec:1stRelaxation},
we compute
the slowest relaxation mode and the mean first passage times
of a KCl nanocluster.
We find there that
the shape of the energy landscape \cite{wales,stillingerBook} tells us
why its relaxation makes effective use of
the shortest routes of the vacancy from the center to the surface.
In Sec.~\ref{sec:2ndRelaxation},
we study the second slowest relaxation mode
and
the corresponding mean first passage times
of the KCl nanocluster.
In Sec.~\ref{sec:symmetric},
we first confirm
that, under exchanging sinks and sources,
the mean first passage time approximation for the relaxation times
in Sec.~\ref{sec:1stRelaxation}
is asymmetric, 
while
that in Sec.~\ref{sec:2ndRelaxation} is suitably symmetric.
We then develop a symmetric stationary population method
for the mean first passage times, 
where they
are interpreted
as the mean first encounter times of particle-hole pairs.
The iterative use of the symmetric method turns out
to be equivalent to an inverse power method for diagonalization of matrices.
We show that
the mean first passage time approximations of the relaxation times
are 
good approximations
that
converge to
the exact 
relaxation times
with the iterative use of the symmetric method.

\section{Markov state model, Relaxation rates, and  Mean first passage times\label{sec:mfpts}}
In this section,
we introduce a Markov state model,
and
we describe
a popular method of calculating first passage times for this model 
according to Refs.~\cite{ngvankampenResub, aldousResub}.
We also show that
the mean first passage times obey stationary population equations,
which will be used to  develop
a symmetrized version of the population method 
later in Sec.~\ref{sec:symmetric}.

\subsection{Continuous-time Markov state model}
We start with a 
continuous-time Markov state model
described by a transition rate matrix $K$
with
finite dimension, $n$, of the state space.
The kinetic equation
is given by
\begin{equation}
\frac{d\vb*{P}}{dt}=K\vb*{P}, 
\label{eq:cmc}
\end{equation}
where 
$\vb*{P}$ 
is the probability distribution $\vb*{P}=(p_1,\dots,p_n)^T$, 
with  $p_i$ denoting  the probability of the $i$th state
and the superscript $T$ denoting the  transpose.
We assume 
that $K$ is time-independent
and satisfies $K_{ij}\geqslant 0$ ($i\neq j$) and
the probability conservation condition of
$\sum_{i=1}^n (K)_{ij}=0$
($j=1,2,\dots,n$).
Further,
we assume
that
the equilibrium, $\lim_{t\to\infty}\vb*{P}(t)$, is a unique vector
$\vb*{P}_\text{eq}$ satisfying the detailed balance conditions
$(K)_{i,j} (\vb*{P}_\text{eq})_j=(K)_{j,i} (\vb*{P}_\text{eq})_i$
\cite{haken, ngvankampenResub}.
Then,
the eigenvalues of $K$ satisfy 
\begin{align}
0= \lambda_0 > \lambda_1 \geqslant \dots \geqslant \lambda_{n-1}. 
\end{align}
The equilibrium $\vb*{P}_\text{eq}$ coincides with
the zeroth eigenvector of $K$,  
and the first, second, $\dots$ eigenvectors $\vb*{P}_i$ of $K$
represent the slowest relaxation modes 
with the relaxation times of $|\lambda_1|^{-1} \geqslant |\lambda_2|^{-1}\geqslant\dots$,
respectively.

\subsection{Mean first passage times}
The mean first passage times, $t_{i,j}$, from a state $j$ to  $i$
are evaluated by  connecting  perfect absorbers to all the final destinations
of
$i$.
The resulting equation
is given by
\begin{equation}
\frac{d\vb*{P}}{dt}=K\vb*{P}-\vb*{S}_-, 
\label{eq:cmcS}
\end{equation}
where $\vb*{S}_-$ represents the perfect absorbers that always keep
$(\vb*{P})_i=0$ for the sink states of $i$.
Without the loss of generality,
the sink states are assumed to be the states of $i=1,\dots,m$,
and
the other  states, which are free from the absorbers, are 
the remainders of  $i=m+1,\dots,n$.
The perfect absorber conditions are represented as follows:
\begin{align}
&\vb*{S}_-=(s_1,\dots,s_m,0,\dots,0)^T\equiv 
\begin{pmatrix}
\vb*{s}_-\\ \vb*{0}
\end{pmatrix},\label{eq:Ss}\\ 
&\vb*{P}=(0,\dots,0,p_{m+1},\dots,p_n)^T\equiv 
\begin{pmatrix}
\vb*{0}\\ \vb*{p}
\end{pmatrix}.
\label{eq:Pp}
\end{align}
By substituting Eqs.~\eqref{eq:Ss}
and \eqref{eq:Pp} for Eq.~\eqref{eq:cmcS},
we have the following solution
with the initial condition of $\vb*{P}_0=
\begin{pmatrix}
\vb*{0}\\
\vb*{p}_0  
\end{pmatrix}
$:
\begin{align}
&\vb*{p}(t)=e^{t K_{FF}}\vb*{p}_0,
\label{eq:kai_p}\\
&\vb*{s}_-(t)=K_{SF}e^{t K_{FF}}\vb*{p}_0,
\label{eq:kai_s}
\end{align}
where
$\vb*{p}_0$ satisfies
\begin{align}
\|\vb*{p}_0\|\equiv \sum_{i=1}^{n-m} |(\vb*{p}_0)_i|
=\sum_{i=1}^{n-m} (\vb*{p}_0)_i
=1.\label{eq:p0norm}
\end{align}
$K_{FF}$
is the submatrix with dimension $(n-m)\times (n-m)$
formed by selecting the rows of $K$ from $m+1$ to $n$ and the columns from $m+1$ to $n$,
and
$K_{SF}$
is
the 
submatrix with dimension $m\times (n-m)$
formed by
selecting the rows from 1 to $m$ and the columns from $m+1$ to $n$.
The probability conservation property of the rate matrix of $K$
can be represented by
\begin{align}
\sum_{i=1}^{m} (K_{SF})_{ij}+\sum_{i=1}^{n-m} (K_{FF})_{ij}=0
\label{eq:pcons}
\end{align}
for $j=1,2,\dots,n-m$.
Multiplying both sides of Eq.~\eqref{eq:pcons}
by $(K_{FF}^{-1})_{jk}$ and adding the resultant equations from $j=1$ to $n-m$,
we have the following equations
\begin{align}
\sum_{i=1}^m (-K_{SF}K_{FF}^{-1})_{ik}=1\quad
(k=1,2,\dots,n-m).
\label{eq:useful}
\end{align}

The $i$th element of $\vb*{s}_{-}(t)$ 
describes
the first passage time distribution
of being absorbed in the $i$th sink
at time $t$.
Therefore,
by integrating Eq.~\eqref{eq:kai_s} from $t=0$ to $\infty$,
the probability of
being absorbed in the $i$th sink for $0\leqslant t<\infty$
is given by the $i$th element of 
\begin{equation}
\bar{\vb*{s}}_{-}
=\int_0^\infty \vb*{s}_{-}(t) dt
=-K_{SF}K_{FF}^{-1} \vb*{p}_0,
\label{eq:sbar-}
\end{equation}
where we use Eq.~\eqref{eq:kai_s} and
$\lim_{t\to\infty}e^{t K_{FF}}=0$,
which  holds because all eigenvalues of $K_{FF}$ are negative values.
From Eq.~\eqref{eq:sbar-}
and
$({\vb*{s}}_{-}(t))_i \geqslant 0$,
we see that $(\bar{\vb*{s}}_{-})_i \geqslant 0$.
Moreover,
with the use of Eqs.~\eqref{eq:p0norm} and \eqref{eq:useful},
we have
\begin{align*}
\|\bar{\vb*{s}}_- \|&\equiv
\sum_{i=1}^m  |(\bar{\vb*{s}}_-)_i |
=\sum_i \left( \bar{\vb*{s}}_{-}\right)_i\\
&=\sum_i \left(-K_{SF}K_{FF}^{-1}\vb*{p}_0\right)_i\\
&=\sum_{i,k}  \left(-K_{SF}K_{FF}^{-1}\right)_{ik}\left(\vb*{p}_0\right)_k \\
&=\sum_{k} \left(\vb*{p}_0\right)_k 
=\|\vb*{p}_0\|=1,
\end{align*}
whence
\begin{align}
\|\bar{\vb*{s}}_-\|
=\|\vb*{p}_0\|=1.
\end{align}

The conditional probability distribution $\rho_i(t)$
of being absorbed at time $t$ when the system is known to be
absorbed in the state of $i$ is given by
\begin{align}
\rho_i(t)=\frac{ \left[  \vb*{s}_{-}(t)\right]_i }
{ \left(  \bar{\vb*{s}}_{-}\right)_i }
=
\frac{(K_{SF} e^{t K_{FF}}\vb*{p}_0)_i}{(-K_{SF}K_{FF}^{-1} \vb*{p}_0)_i}. 
\end{align}
Therefore,
the mean first passage time, $t_{i,j}$,
from the state $j$ to the sink state $i$
is 
given by
\begin{align}
t_{i,j}
&=\int_0^\infty t \rho_i(t)dt\nonumber\\
&=
\frac{(K_{SF}K_{FF}^{-2} \vb*{p}_0)_i}
{(-K_{SF}K_{FF}^{-1} \vb*{p}_0)_i}
\label{eq:tij}
\end{align}
with  $(\vb*{p}_0)_k=\delta_{k,j-m}$ ($k=1,2,\dots,n-m$).
Moreover,
the mean first passage time, $t_j$,
from the state $j$ to any absorbing states
is 
given by
\begin{align*}
t_j &=\sum_i (\bar{\vb*{s}}_-)_i t_{i,j}
=\sum_i \left(K_{SF}K_{FF}^{-2}\vb*{p}_0\right)_i\nonumber\\ 
&=\sum_{i,k}  \left(-K_{SF}K_{FF}^{-1}\right)_{ik}\left(-K_{FF}^{-1}\vb*{p}_0\right)_k\nonumber\\
&=\sum_{k} \left(-K_{FF}^{-1}\vb*{p}_0\right)_k, 
\end{align*}
that is
\begin{align}
t_{j}
&= \| -K_{FF}^{-1} \vb*{p}_0\|,
\label{eq:allsinks}
\end{align}
where
Eqs.~\eqref{eq:useful},
\eqref{eq:sbar-},
and
\eqref{eq:tij}
are used.

Equations (\ref{eq:tij})
and 
(\ref{eq:allsinks})
are the basic formulas
for 
calculating
the mean first passage times.

Next,
we show that
the mean first passage times 
can be evaluated
from a stationary population distribution.
Let us consider
the following mean residence time distribution,
\begin{equation}
\bar{\vb*{p}}
=\int_0^\infty \vb*{p}(t)dt=-K_{FF}^{-1}\vb*{p}_0,
\label{eq:pbar}
\end{equation}
where
$( \bar{\vb*{p}})_i$ 
is  the mean residence time in the $(i+m)$th state
for $i=1,\dots,n-m$.
Hence,
the mean residence time in the whole system
is given by the sum, $\|\bar{\vb*{p}}\|$, of $( \bar{\vb*{p}})_i$ from $i=1$ to $n-m$,
which is, of course,  equivalent to 
the mean first passage time $t_j$ of  Eq.~(\ref{eq:allsinks}).

Equation \eqref{eq:pbar} enables us to confirm that
$\bar{\vb*{p}}$
satisfies the following non-equilibrium stationary state equation:
\begin{align}
\frac{d\bar{\vb*{p}}}{dt}
=K_{FF}\bar{\vb*{p}}+\bar{\vb*{s}}_+
=\vb*{0}
\label{eq:stationaryEq}
\end{align}
with $\bar{\vb*{s}}_+=\vb*{p}_0$.
Hence,
we can interpret
$\bar{\vb*{s}}_+$
as
the source term 
that adds one particle with distribution $\vb*{p}_0$ per unit time,
$\bar{\vb*{p}} $ as the stationary population of Eq.~\eqref{eq:stationaryEq}, 
$\|\bar{\vb*{p}}\|$ as
the total population contained in $\bar{\vb*{p}} $,
and
$\|\bar{\vb*{p}}\|^{-1}$ as
the probabilistic flow  carried by one particle.
Namely,
we can also compute the mean first passage times
as
the total numbers, $\|\bar{\vb*{P}}\|$, of particles 
in the stationary
population  $\bar{\vb*{P}} $
obeying
the following stationary equation: 
\begin{align}
\frac{d}{dt} \bar{\vb*{P}}
=K \bar{\vb*{P}}+\vb*{S}_+ - \vb*{S}_-
=\vb*{0},
\label{eq:population}
\end{align}
where
\begin{align}
\bar{\vb*{P}}=
\begin{pmatrix}
\vb*{0}\\
\bar{\vb*{p}}
\end{pmatrix},\quad
\vb*{S}_+=
\begin{pmatrix}
\vb*{0}\\
\vb*{p}_0
\end{pmatrix},\quad
\vb*{S}_-=
\begin{pmatrix}
\bar{\vb*{s}}_-\\
\vb*{0}
\end{pmatrix}.
\label{eq:population2}
\end{align}
Note that
the stationary population equation \eqref{eq:population}
will be used in Sec.~\ref{sec:symmetric}.

\section{KCl nanocluster vacancy diffusion model\label{sec:model}}
In this section,
according to Ref.~\cite{niiyama},
we first present
the vacancy diffusion model of KCl nanoclusters 
as an example of a  practical problem,
and
then
we introduce
the corresponding Markov state model of the vacancy diffusion.

\subsection{Local minima and saddle points on the potential energy surface of a KCl nanocluster}
Let us assume that
one chlorine ion is extracted from a cube of ionic crystal
with equal $N_L$-atom edges
and 
further that
$N_L$ is an odd number $2 n_L+1$,
and 
the resultant 
cluster with $N\equiv{N_L}^3-1$ atoms
is electrically neutral.
We employ the two-body Coulomb plus Born-Mayer type potential model,
\begin{align}
v(r_{ij})=
\frac{Q_i Q_j}{4 \pi \epsilon_0 r_{ij}}
+A_{ij} \exp\left( \frac{R_i+R_j-r_{ij}}{\rho}\right),
\end{align}
where
$Q_i$, $Q_j$ are the charges of the $i$th and $j$th atoms,
$\epsilon_0$ is the vacuum permittivity, and
$r_{ij}$ is the distance between the $i$th and $j$th atoms.
We use
the values of the three parameters $A_{ij}$,  $R_i$, and $\rho$
that were introduced by Tosi and Fumi in Ref.~\cite{tosi}:   
$A_{ij}=0.2210$, $0.2637$, and $0.1582\,$ eV, respectively,
for
K--Cl,
K--K,
and Cl--Cl
pairs;
$R_i=1.463$ and $1.585\, \mathrm{\AA}$
for
K and Cl, respectively;
and
$\rho=0.337\, \mathrm{\AA}$.
Then,
the total potential energy of the cluster is given by
\begin{align}
V(\vb*{r}_1,\dots,\vb*{r}_N)=\sum_{i=1}^{N-1} \sum_{j=i+1}^N v(r_{ij}).
\label{eq:totalPot}
\end{align}

In the course of the time evolution,
the vacancy moves around the cluster,
which
introduces  atomic mixing to the cluster.
Note that
the cubic form  of the cluster is kept with the time evolution
when the temperature is sufficiently low \cite{cubic}.
At such low temperatures,
the position of the vacancy is specified
by the cubic lattice point $\vb*{n}=(n_x,n_y,n_z)$ with 
$-n_L \leqslant n_x,n_y,n_z \leqslant n_L$.
Moreover,
we are able to find the atomic structure corresponding to the
vacancy lattice point $\vb*{n}$
as follows:
First,
atoms are arranged at
$d(m_x,m_y,m_z)$ with lattice constant $d=3.147\ {\rm\AA}$ for KCl, 
where $(m_x,m_y,m_z)\neq \vb*{n}$ and $-n_L \leqslant m_x,m_y,m_z \leqslant n_L$.
Then, the  configuration of the atoms
is relaxed to the local minimum (LM)
configuration, $\vb*{r}=(\vb*{r}_1,\dots,\vb*{r}_N)$,
of the potential energy surface, e.g.,
by
the conjugate gradient method \cite{matrixcomp}.
In this way,
$\vb*{n}$ is assigned to
the LM atomic structure as $\vb*{r}_{\vb*{n}}=\vb*{r}$.
We compute the LM configurations $\vb*{r}_{\vb*{n}}$
and the energies $V(\vb*{r}_{\vb*{n}})$
for all $\vb*{n}$. 
The LM datasets of $V(\vb*{r}_{\vb*{n}})$ and $\vb*{r}_{\vb*{n}}$
are stored in a file in nondecreasing order of energy $V(\vb*{r}_{\vb*{n}})$.
For the sake of notational simplicity,
the $i$th lowest energy is denoted as $E_i$,
and
the corresponding
LM, atomic configuration, and
vacancy lattice point 
are denoted as
$i$,
$\vb*{r}_i$, and
$\vb*{n}_i$,
respectively.
Then, 
we proceed to find out all of the saddle points (SPs) connecting the adjacent
LMs,
e.g.,
by
the nudged elastic band method \cite{wales}.
The corresponding saddle point connecting the $i$th and $j$th LMs,
atomic configuration, and potential energy
are denoted as 
${ij}$,
$\vb*{r}_{ij}$, and
$E_{ij}$,
respectively.
For the computational details of enumerating all of the LMs and SPs,
we refer the reader to Ref.~\cite{niiyama}.

\subsection{Markov state model of KCl vacancy diffusion}
Let $f(\vb*{r})$ denote the  probability density function at 
a configuration $\vb*{r}$. 
We suppose
that
the intra-LM relaxations are
so fast
that
$f(\vb*{r})$ is represented as
\begin{align}
f(\vb*{r})= p_1 f_1(\vb*{r})+p_2 f_2(\vb*{r})+\dots+p_n f_n(\vb*{r}),
\end{align}
where
$n$ is  the number of the LMs,
$f_i(\vb*{r})$ is
the local equilibrium in the $i$th-LM basin,
and
$p_i$ is the probability that
$\vb*{r}$ is in the $i$th-LM basin.
We identify 
$p_i$ 
with the probability in  the $i$th state of the Markov state model.
Then,
the probability vector $\vb*{P}$ of the Markov state model
is given by
$\vb*{P}=(p_1,p_2,\dots,p_n)^T$.

Next,
we  evaluate
the transition rate $k_{i,j}$
from  the $j$th to the adjacent $i$th state, 
when 
the potential barrier energies are sufficiently larger than 
the average kinetic energy of $k_\text{B} T/2$ for one degree of freedom
at temperature $T$,
where $k_{\mathrm{B}}$ denotes the Boltzmann constant.
In this case,
the transition rate $k_{i,j}$
from  the $j$th to the $i$th state 
is given by
\begin{align}
k_{i,j}=
\nu_{i,j}
\exp \left({-\frac{E_{ij}-E_j}{k_\mathrm{B} T}}
\right).
\label{eq:kij}
\end{align}
Here,
the prefactor $\nu_{i,j}$,
called  a frequency factor,
is given by
\begin{align}
\nu_{i,j}=\frac{\prod_k^\prime (\vb*{\nu}_i)_k}{\prod_k^\prime (\vb*{\nu}_{ij})_k},
\label{eq:nuij}
\end{align}
where $\vb*{\nu}_i$ and $\vb*{\nu}_{ij}$ 
are
vibrational frequency vectors
that are calculated from 
the Hessians
at
$\vb*{r}_i$ and $\vb*{r}_{ij}$, respectively.
The product
$\prod^\prime_k (\vb*{\nu}_\ast)_k$ denotes
the partial product of the positive frequency modes
$(\vb*{\nu}_\ast)_k >0$,
where
the imaginary frequency modes and
the zero frequency modes
are left out from the products.

Finally,
the transition rate matrix $K$ is
given by
\begin{align}
(K)_{i,j}=k_{i,j}\ (i\neq j)
\quad \text{and}\quad
(K)_{j,j}=-\sum_{i\neq j} k_{i,j},\label{eq:Kmat}
\end{align}
where
the probability conservation equations
$\sum_{i}(K)_{i,j}=0$ for $j=1,2,\dots,n$
and
the detailed balance conditions
$(K)_{i,j} (\vb*{P}_\text{eq})_j=(K)_{j,i} (\vb*{P}_\text{eq})_i$ 
for $i$, $j=1,2,\dots,n$
are satisfied
since $(\vb*{P}_\text{eq})_i\propto \exp(-E_i/k_\text{B}T)$.

In the following,
we examine the slowest relaxation modes
of the  KCl vacancy diffusion model of $N_L=13$.
To this end,
we  searched for the LMs and the SPs of the cluster,
thereby finding  $1099$ ($=1+(N_L^3-1)/2$) LMs and 5472 SPs.
Then,
with the use of Eqs.~\eqref{eq:kij}, \eqref{eq:nuij}, and \eqref{eq:Kmat},
we formed its rate matrix of $K$ at $k_\text{B} T=0.03$ eV,
whose  matrix dimension $n$ is $1099$
and
the number of nonzero offdiagonal elements
is
$10948$ ($=5472\times 2$).

By diagonalizing $K$, 
we obtained the eigenvalues $\lambda_i$ and 
the corresponding relaxation modes $\vb*{P}_i$
for $i=0,1,2,\dots, n-1$,
where
$\vb*{P}_0=\vb*{P}_\text{eq}$.
In Sec.~\ref{sec:1stRelaxation} and Sec.~\ref{sec:2ndRelaxation},
we study the properties of
the slowest relaxation $\vb*{P}_1$
and the second slowest relaxations $\vb*{P}_2$,
$\vb*{P}_3$, and $\vb*{P}_4$ ($\lambda_2=\lambda_3=\lambda_4$),
respectively.

\section{the slowest relaxation mode\label{sec:1stRelaxation}}
In this section,
we show that
the slowest relaxation mode of the KCl nanocluster 
makes  effective use of the fast surface diffusion of the cluster,
in terms of
the mean first passage times,
the free energy landscapes,
and
the atomic interactions.

\subsection{Dominant pathways\label{sec:dominantPathways}}

\begin{figure}
\begin{center}
\includegraphics[width=6.5cm]{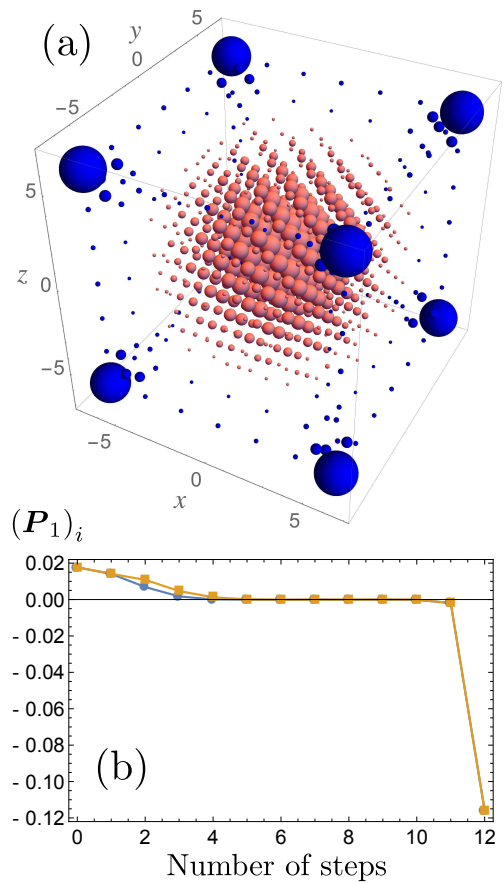} 
\end{center}
\caption{%
The slowest relaxation mode, $\vb*{P}_1$, at  $k_\text{B}T=0.03$ eV:
(a)
 Red (light gray) and blue (dark gray) balls  are depicted
 at vacancy lattice points of $\vb*{n}_i$ 
 with radii $\propto |(\vb*{P}_1)_i|^{1/3}$
for $(\vb*{P}_1)_i > 0$  and $(\vb*{P}_1)_i < 0$,
respectively.
Note that
$\vb*{P}_1$ has  cubic symmetry around the $x$-, $y$-, and $z$-axes.
 (b)
 $(\vb*{P}_1)_i$ along pathways $\ell_1$ of Eq.~\eqref{eq:ell1}
 and $\ell_2$ of Eq.~\eqref{eq:ell2}
 are plotted as a function of
 the number of steps from the origin $(0,0,0)$
 with blue circles connected by the lower line
 and orange squares connected by the upper line, respectively.
}
\label{fig:density}
\end{figure}

\begin{figure}[t]
\begin{center}
\includegraphics[height=6.5cm]{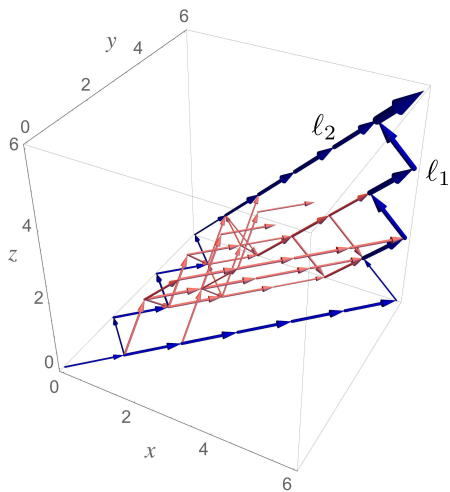} 
\end{center}
\caption{%
Probabilistic flows $f_{i,j}$ of $\vb*{P}_1$
at  $k_\text{B}T=0.03$ eV
are represented
by
arrows from  $\vb*{n}_j$ 
to $\vb*{n}_i$ 
with  cylinder radii $\propto\sqrt{|f_{i,j}|}$
for $f_{i,j}>0$. 
The flows have the same cubic symmetry as 
in  Fig.~\ref{fig:density}.
Hence,
only the flows in a reduced zone $n_y\geqslant n_x \geqslant n_z \geqslant 0$
are represented.
We see that  two  pathways $\ell_1$ and $\ell_2$ in blue (dark gray)
carry the dominant flows:
 $\ell_1$ is composed of
the 
straight move from the origin $(0,0,0)$ to the edge center $(6,6,0)$
and 
 the succeeding zigzag move to the vertex $(6,6,6)$;
$\ell_2$ is composed of 
the zigzag move from the origin to the face center  $(0,6,0)$
 and the succeeding straight move to the vertex.
 [See Eqs.~\eqref{eq:ell1} and \eqref{eq:ell2}.]
}
\label{fig:flow}
\end{figure}

Figure~\ref{fig:density}(a)
shows the slowest relaxation mode of $\vb*{P}_1$,
from which
we see
that
$\vb*{P}_1$ has
the probability excesses at around the origin $(0,0,0)$
and the probability shortages at around
the eight vertices of $(\pm n_L,\pm n_L,\pm n_L)$.
In Fig.~\ref{fig:density}(b),
we also
plot
the values of
$(\vb*{P}_1)_i$ along two pathways
from the origin to the vertex $(n_L,n_L,n_L)$,
which clearly shows
that
they have the maximum values at the origin
and
 positive values up to three steps from the origin
 and negative values at the vertices.

$\vb*{P}_1$ decays with the rate of
$\lambda_1=-1.92 \times 10^{5}\ \mathrm{s}^{-1}$ 
over the course of time.
Hence,
the probabilistic flow
from the center to the vertices
is expected in the relaxation process of $\vb*{P}_1$.
To confirm this,
we compute all of 
the probabilistic flows $f_{i,j}$ from $j$ to $i$,
generated by $\vb*{P}_1$,
where
$f_{i,j}$ is 
given by
$f_{i,j}=k_{i,j}p_j-k_{j,i}p_i$
with $p_i=(\vb*{P}_1)_i$
for $1\leqslant i, j \leqslant n$.
In Fig.~\ref{fig:flow},
$f_{i,j}$  are  represented 
by the arrows from $\vb*{n}_j$ to $\vb*{n}_i$ when  $f_{i,j} > 0$.
We see that
the probabilistic flows from the center to the vertices
are generated.
More precisely,
the probabilistic flows are not uniform
but mostly along the two dominant pathways of
$\ell_1$ and $\ell_2$ as depicted in Fig.~\ref{fig:flow}.

The dominant pathways of $\ell_1$ and $\ell_2$
are defined 
by the following  algorithm that searches for the maximum flow
pathways flowing into  the terminals.
First,
we start with
the  terminal of
the vertex $(n_L,n_L,n_L)=(6,6,6)$.
The probabilistic flow from $(5,6,5)$ 
flows into the terminal $(6,6,6)$.
The probabilistic flows from
$(6,6,5)$ and $(4,6,4)$
flow into $(5,6,5)$, respectively.
The dominant pathways via
$(6,6,5)$ and $(4,6,4)$
are denoted as $\ell_1$ and $\ell_2$,
respectively.
We then search for the source flow of $\ell_1$ as follows.
The probabilistic flow from $(5,6,4)$ 
is the maximum flow flowing into  $(6,6,5)$.
That from $(6,6,3)$ 
is the maximum flow flowing into  $(5,6,4)$,
and so on.
This procedure continues until the source $(0,0,0)$ appears
and 
gives the dominant pathway as
\begin{align}
\ell_1
=&
(0, 0, 0)
\to (1, 1, 0)
\to \dots \to (6, 6, 0)\nonumber\\
&\to (5, 6, 1)
\to (6, 6, 2)
\to (5, 6, 3)\label{eq:ell1}\\
&\to (6, 6, 4)
\to (5, 6, 5)
\to (6, 6, 6). \nonumber
\end{align}
The dominant paths of this kind  are composed of six straight steps
from the origin to the 12 centers of the edges
$(\pm 6,\pm 6,0)$,
$(\pm 6,0,\pm 6)$,
$(0,\pm 6,\pm 6)$,
followed by six zigzag steps from there to the vertices along the edges.
Similarly,
we  search for the source flows flowing into $(4,6,4)$
and obtain $\ell_2$ as
\begin{align}
\ell_2=
& (0,0,0)
\to (1,1,0)
\to (0, 2,0)
\to (1, 3,0)\nonumber\\
&\to (0, 4,0)
\to (1, 5,0)
\to (0, 6,0)\to\label{eq:ell2}\\
&
(1, 6,1)\to
\dots
\to (5, 6,5)
\to (6, 6, 6). \nonumber
\end{align}
The dominant paths of the second kind are composed of six zigzag steps
from the center to the six centers of the faces,
$(\pm 6,0,0)$,
$(0,\pm 6,0)$,
$(0,0,\pm 6)$,
followed by six straight steps from there to the vertices.

In other words,
the dominant paths arriving at each vertex
are the three 
$\ell_1$-type paths,
which climb along the three edges connected to the vertex,
and the three 
$\ell_2$-type paths,
which move across the three faces containing the vertex.
Note that
these observations
are consistent with our previous results from Ref.~\cite{rg}.
There,
all states are
divided into groups, called metabasins, 
that are located  around
the vertices,
the edges,
the faces,
and the center part,
and then
the relaxation processes
are 
described accurately
by
the renormalized transitions
between 
these metabasins.
That is to say,
we have reconfirmed here that
the essential pathways
connecting
the vertices,
the edges,
the faces,
and the center part
are indispensable for
describing the slowest transport of probabilities.

\subsection{Mean first passage times\label{sec:mfpt2vertex}}
\begin{figure}[t]
\begin{center}
\includegraphics[width=7cm]{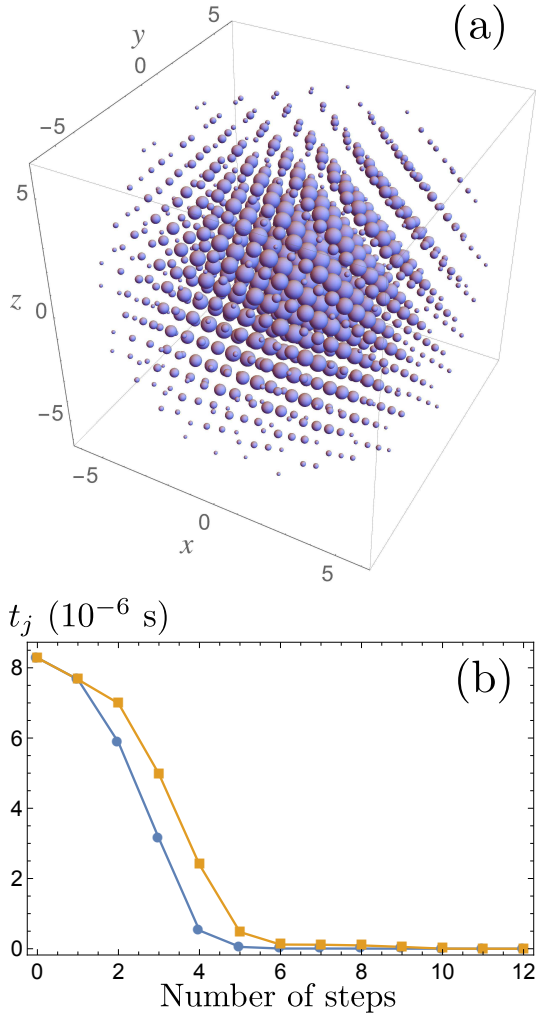}
\end{center}
\caption{%
Mean first passage times of $t_j$  at $k_\text{B}T=0.03$ eV,
with perfect sinks connected to the vertices $(\pm 6,\pm 6,\pm 6)$:
(a) $t_j$ are represented
by balls of radii $\propto |t_j|^{1/3}$ located at $\vb*{n}_j$. 
The  mean first passage times
 have cubic symmetry around the $x$-, $y$-, and $z$-axes.
 (b)  $t_j$  along $\ell_1$ and $\ell_2$
 are plotted as a function of the number of steps from the origin
 with blue circles connected by the lower line
 and orange squares connected by the upper line, respectively.
Both of $t_j$ have  the maximum value of $8.28\times 10^{-6}$ s
 at the origin of $(0,0,0)$.
 The values of $t_j$ in the center part,
 where the number of steps is from 0 to 4,
have the same order of magnitude,
while
 those in the two layers from  the surface,
 where the number of steps is from 5 to 12,
 are quite small values.
}
\label{fig:mfpts}
\end{figure}

Here,
we consider why
the dominant paths carrying large probabilistic flows
are not
almost straight, nine-step shortest paths from the origin to the vertices,
such as
$
(0, 0, 0)
\to (0, 1, 1)
\to (1, 1, 2)
\to (2, 2, 2)
\to (2, 3, 3)
\to (2, 4, 4)
\to (3, 4, 5)
\to (4, 5, 5) 
\to (5, 5, 6) 
\to (6, 6, 6)
$,
but longer 12-step paths of $\ell_1$ and $\ell_2$  in Fig.~\ref{fig:flow}.
To this end,
we examine the mean first passage times of $t_j$ from
various initial states of $j$
to the sink states of the vertices $(\pm n_L,\pm n_L,\pm n_L)$.

Using Eq.~\eqref{eq:allsinks},
we compute $t_j$ for  various initial states of $j$.
The resulting $t_j$
are
plotted in Fig.~\ref{fig:mfpts}(a).
We see that
the states in the central part have large $t_j$ values,
while
the states on the surface have quite small values.
That is,
the KCl nanocluster is a hybrid system
that combines entirely different microscopic diffusive regions:
The central part is the region that is hard to move stochastically,
whereas
the surface part is the region that is quite easy to move.
To see this more closely,
we plot $t_j$ along the paths of $\ell_1$ and $\ell_2$
in Fig.~\ref{fig:mfpts}(b),
where
both of $t_j$ have the maximum value at the origin
and
they are negligibly small compared to the maximum value
in two layers from the surface.

Now,
we see the reason  why the detoured pathways are selected
to be the dominant pathways,
as depicted in Fig.~\ref{fig:flow}.
Namely,
it is because
all the dominant pathways prefer to pass
the slow diffusion region of the central part as soon as possible,
with the fewest steps of $n_L=6$,
in order to make the most effective use of the fast diffusion in the surface region.

Next,
we show that
$\lambda_1$
can be evaluated approximately from  $t_j$.
The longest  mean first passage time to the vertices
is
$t_{(0,0,0)}=8.28\times10^{-6}$ s.
The probability of being at the vertices in equilibrium
is  $\vb*{P}_\text{eq}(\text{vertices})=
\sum_{i \in \text{vertices}} (\vb*{P}_\text{eq})_i
=0.885$.
We regard
the equilibration time
as
the required time of
constructing $\vb*{P}_\text{eq}(\text{vertices})$.
Then,
the equilibration time
is approximately given by
\begin{align}
\vb*{P}_\text{eq}(\text{vertices})\times t_{(0,0,0)}
=7.33\times 10^{-6} \ \text{s}.
\label{eq:P1relaxtime}
\end{align}
The corresponding equilibration rate
is given by the inverse of the equilibration time,
$1.36\times10^{5}\ \mathrm{s}^{-1}$.
The estimate
agrees 
qualitatively 
with the values of $|\lambda_1|=1.92\times 10^5$ s$^{-1}$,
although
it is a  smaller value  than $|\lambda_1|$.

This discrepancy arises because,
although the actual excess probabilities in $\vb*{P}_1$
are distributed in the central part
as depicted in Fig.~\ref{fig:density},
the excess probability is approximated to
the distribution concentrated on 
the origin, for the mean first passage time approximation of $\lambda_1$.
We will revisit this point
in Sec.~\ref{sec:symmetric}.

\subsection{Free energy sequences\label{sec:freeEnergySeq}}

\begin{figure}[t]
\begin{center}
\includegraphics[width=7.5cm]{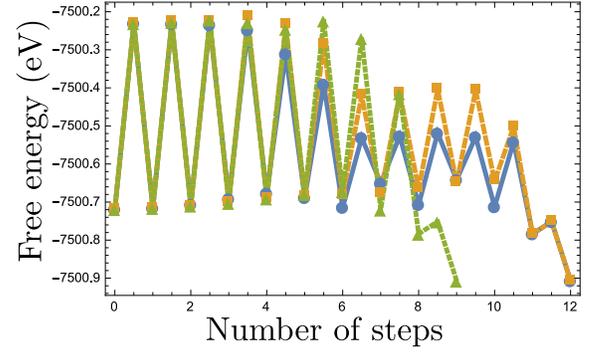} 
\end{center}
\caption{%
Free energy sequences in units of eV
are plotted
at  $k_\text{B}T=0.03$ eV,
as functions of steps counted from the origin $(0,0,0)$ to a vertex
along the geometric shortest path
({$\blacktriangle$}, green),
 along the dominant path, $\ell_1$, of Eq.~\eqref{eq:ell1}
 via an edge center
(\raisebox{-0.25ex}{\Large $\bullet$}, blue)
and the dominant path, $\ell_2$, of Eq.~\eqref{eq:ell2} via a  face center ($\blacksquare$, orange).
The integer steps
indicate the free energies of the local minima,
and
the half-integer steps 
indicate the free energies of the saddle points
that connect the basins of adjacent local minima.
}
\label{fig:FijFj}
\end{figure}

Here,
we consider the physical reason why 
the  bottleneck of diffusion in the Markov state model of the KCl nanocluster
is located at the central part.
To this end,
we examine the following free energies
for the LMs of $j$ and  the SPs of $ij$,
respectively:
\begin{align}
 F_{j}&=E_j-k_\text{B}T \ln {\prod_k}^\prime (\vb*{\nu}_j)_k,\label{eq:Fj}\\
 F_{ij}&=E_{ij}-k_\text{B}T \ln {\prod_k}^\prime (\vb*{\nu}_{ij})_k,
 \label{eq:Fij}
\end{align}
where  $k_{i,j}=\exp[  -\beta(F_{ij}-F_j) ]$ holds.
Then,
the free energy sequence of local minima and saddle points
along a pathway $i_0\to i_1\to \cdots \to i_s$
is given by
$F_{i_0},F_{i_0i_1},F_{i_1},F_{i_1i_2},F_{i_2},\dots F_{i_s}$.

In Fig.~\ref{fig:FijFj},
we plot the free energy sequences of 
the dominant pathways of $\ell_1$ [Eq.~\eqref{eq:ell1}]
and $\ell_2$ [Eq.~\eqref{eq:ell2}].
Along these dominant pathways,
the activation energies for the inner transitions $i_k\to i_{k+1}$
are about
$\Delta F_{i_{k+1},i_k}=F_{i_ki_{k+1}}-F_{i_k}\approx 0.5$ eV,
and hence
the transition rates become quite low rates of
$k_{i_{k+1},i_k}\approx 6\times 10^{5}\ \mathrm{s}^{-1}$
at $k_\text{B}T=0.03$ eV.
In contrast,
those for the surface transitions
are
$\Delta F_{i_{k+1},i_k}\approx 0.2$ eV,
and
the transition rates are about
$k_{i_{k+1},i_k}\approx 1\times 10^{10}\ \mathrm{s}^{-1}$,
which are about $10^4$ times higher than the inner rates,
at the same temperature.

For comparison,
we also plot 
the free energy sequence 
along 
the nine-step geometric shortest path
in Fig.~\ref{fig:FijFj}.
We see that
the first seven steps are 
in the slow diffusion  region
and 
the last two steps are 
in the faster surface diffusion region.
Therefore,
the geometric shortest path
cannot be dominant,
because 
the extra steps in the slower diffusion region reduce 
its diffusive flow drastically.

Note that
the activation free energies of $\Delta F_{i_k,i_{k+1}} \gtrsim 0.2$ eV
are sufficiently larger than $k_\text{B} T=0.03$ eV
and hence 
the harmonic approximation \eqref{eq:kij}
used in this study is accurate.

\begin{figure}[t]
\begin{center}
\includegraphics[width=5cm]{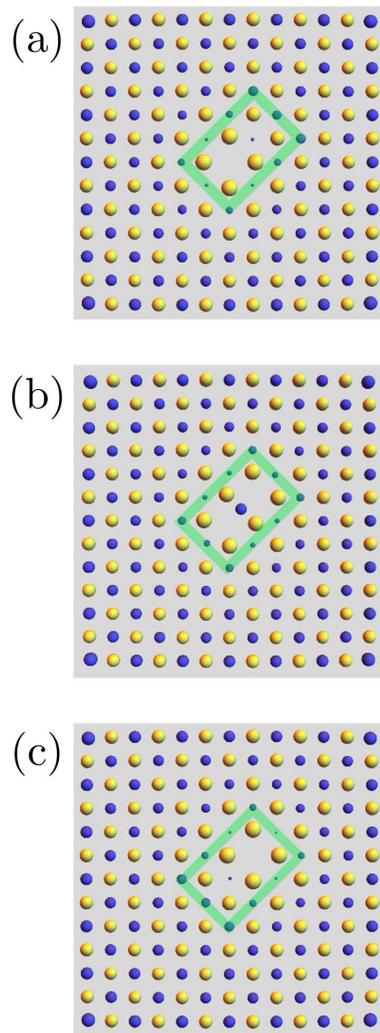} 
\end{center}
\caption{%
Individual energies $V_k$ of $k$th atoms
contained in the $z=0$ planes are shown for
(a)
the LM of the vacancy lattice point of  $\vb*{n}_1=(0,0,0)$,
(b) the SP connecting LMs of $\vb*{n}_1$ and $\vb*{n}_2=(1,1,0)$,
 and
 (c) the LM of  the vacancy lattice point of $\vb*{n}_2$.
The Cl and K atoms of individual energies $V_k$ are
represented by blue (dark gray) and yellow (light gray) balls with radii 
$\propto \sqrt[3]{V_k-\min_{k}\{V_k\} }$.
The defect neighbors are defined by the regions
inside the green (gray) frames of  $|x-y|<1.2d$, $-1.2d<x+y<3.2d$,
 and $|z|<2.2$. (See the text.)
}
\label{fig:LM1SPLM2}
\end{figure}

\subsection{Activation energies on potential energy landscapes\label{sec:activation}}
Next,
we show that
it
can be understood
in terms of  the interatomic interaction energies
why
the surface activation free energies
are so small compared to the inner ones.

First,
from Eqs.~\eqref{eq:nuij}, \eqref{eq:Fj}, and \eqref{eq:Fij},
$\Delta F_{i,j}=  E_{ij}-E_j+k_{\mathrm{B}}T \ln \nu_{i,j}$ holds.
At the low temperature of  $k_\text{B}T=0.03$ eV,
$k_{\mathrm{B}}T \ln \nu_{i,j} \approx 0.03 $ eV
 is negligible compared to $\Delta F_{i,j} \approx 0.5$ eV,
and
 $\Delta F_{i,j} \approx E_{ij}-E_j$ holds.
 Hence,
in the following
we 
consider the activation energies,
$\Delta E_{i,j}=E_{ij}-E_j$,
of  various transitions.

We examine the activation energy of $\Delta E_{i,j}$
when the vacancy lattice points $\vb*{n}_i$ and $\vb*{n}_j$
are present in the inner part of the cluster.
In this case,
$E_i$ and  $E_j$ are almost the same,
as shown in Fig.~\ref{fig:FijFj},
and hence
$\Delta E_{i,j}$ is  determined by
the energy increase from $E_i\simeq E_j$,
due to
the deformation of the crystal structure near the lattice defect.
To quantify the deformation energies,
we introduce the individual potential energy of the $k$th atom as
\begin{align}
V_k=\frac12\sum_{l\neq k} v(r_{kl}).
\end{align}
Then, the total potential energy of Eq.~\eqref{eq:totalPot}
is represented as
\begin{align}
V(\vb*{r})=\sum_{k}V_k. 
\end{align}
Note that
$V_k$ is half of the required energy to remove the $k$th atom from the cluster,
since
$V(\vb*{r}_1,\dots,\vb*{r}_k,\dots\vb*{r}_N)-V(\vb*{r}_1,\dots,\vb*{r}_{k-1},\vb*{r}_{k+1}\dots,\vb*{r}_N)=2 V_k$ holds.
Figures \ref{fig:LM1SPLM2}(a) and \ref{fig:LM1SPLM2}(c), respectively,
show the values of $V_k$ for 
the LMs of the vacancy lattice points of
$\vb*{n}_1={(0,0,0)}$ and  $\vb*{n}_2={(1,1,0)}$.
We see that
the changes of $V_k$ are concentrated
in the vicinities of the vacancies
at around  $\vb*{r}_1= d \vb*{n}_1$ and $\vb*{r}_2= d \vb*{n}_2$
with lattice constant $d$,
and that
$V_k$ of Cl and K atoms
decrease and increase, respectively,
when approaching the vacancy positions.
Figure \ref{fig:LM1SPLM2}(b)
shows the values of $V_k$
for the SP connecting the LMs of $\vb*{n}_1$ and  $\vb*{n}_2$.
The SP has 
the high energy Cl atom as the lattice defect
at around the midpoint, $\vb*{r}_{1,2}=d(1/2,1/2,0)$, of the vacancy positions.
In this case, too,
$V_k$ of Cl and K atoms
decrease and increase, respectively,
when approaching the defect of the high-energy Cl atom.

Next,
we show that
$\Delta E_{i,j}$ can be estimated
with use of the local $V_k$ values around the defects.
To this end,
we obtain the local energies 
$E_1^{\text{loc.}}=-134.34$ eV,
$E_{1,2}^{\text{loc.}}=-133.804$ eV, and
$E_2^{\text{loc.}}=-134.334$ eV,
which
are
the sums of  $V_k$ inside the local regions
surrounded by the green (gray) rectangle frames
in Figs.~\ref{fig:LM1SPLM2}(a), \ref{fig:LM1SPLM2}(b), and \ref{fig:LM1SPLM2}(c),
respectively.
Hence,
the activation energies
evaluated from these local energies 
are given by
$\Delta E_{2,1}^{\text{loc.}}=E_{1,2}^{\text{loc.}}-E_{1}^{\text{loc.}}=
0.530$ eV
and 
$\Delta E_{1,2}^{\text{loc.}}=E_{1,2}^{\text{loc.}}-E_{2}^{\text{loc.}}
=0.536$ eV,
which agree qualitatively with
the exact activation energies
of 
$\Delta E_{2,1}=0.58$ eV and
$\Delta E_{1,2}=0.58$ eV.

Similarly,
we also evaluate the local activation energies of
other types of activation processes
as listed in Table~\ref{table:deltaE}.
We see that
the other types of activation energies
are also  described suitably
by the local activation energies.
Therefore,
we have confirmed that
all of the activation energies can be interpreted as
the energy rises due to the local lattice deformations
generated around the lattice defects.

Also,
the local deformation assumption leads to
the  approximate relations of
$\Delta E_{\text{face}\leftarrow\text{face}}\approx
\Delta E_{\text{inner}\leftarrow\text{inner}}/2$
and
$\Delta E_{\text{edge}\leftarrow\text{face}}\approx
\Delta E_{\text{inner}\leftarrow\text{inner}}/4$,
which are implied in Table \ref{table:deltaE}.
\begin{table*}[t]
\caption{%
Activation energies $\Delta E_{i,j}$
and local activation energies $\Delta E^{\mathrm{loc.}}_{i,j}$
of vacancy transitions in  the KCl cluster of $n_L=6$
 are enumerated  in units of eV.
 $\Delta E_{1,2}=E_{1,2}-E_2$
 and
 $\Delta E^\text{loc.}_{1,2}=E^\text{loc.}_{1,2}-E^\text{loc.}_2$,
 where
 $E^\text{loc.}_{1,2}$ and $E^\text{loc.}_2$ are
  sums of individual atomic energies around the defect points. 
(See the text.)
}
\label{table:deltaE}
\centering
\begin{tabular}{p{3.5cm}p{3.5cm}p{3.5cm}p{3.5cm}}
\hline
\hfil Activation type \hfil  & \hfil $\vb*{n}_1\leftarrow \vb*{n}_2$\hfil  & \hfil $\Delta E_{1,2}$\hfil  &\hfil $\Delta E_{1,2}^{\mathrm{loc.}}$\hfil\\
\hline \hline
\hfil$\text{Inner}\leftarrow\text{Inner}$\hfil  &\hfil $(1,1,0)\leftarrow (0,0,0)$\hfil  &\hfil $0.58$\hfil &\hfil$0.53$ \hfil\\
\hfil$\text{Face}\leftarrow\text{Face}$\hfil  & \hfil$(6,1,1)\leftarrow(6,0,0)$\hfil   &\hfil $0.32$\hfil &\hfil $0.33$\hfil \\
\hfil $\text{Edge}\leftarrow\text{Face}$
\hfil  & \hfil$(6,6,0)\leftarrow (6,5,1)$\hfil  & \hfil $0.15$\hfil &\hfil $0.18$ \hfil\\
\hfil $\text{Vertex}\leftarrow\text{Face}$
\hfil  & \hfil$(6,6,6)\leftarrow (6,5,5)$\hfil  & \hfil $0.027$\hfil &\hfil $0.036$ \hfil\\
\hline
\end{tabular}
\end{table*}
To understand these relations, 
we assume for simplicity that the deformation energy is uniformly distributed 
inside the ball of radius $a \approx d$ located at the defect point.
Assuming further
that
the deformation energy per unit volume is given by  $\epsilon$,
then the activation  energies for inner vacancies are
estimated as
$\Delta E_{\text{inner}\leftarrow\text{inner} }=4\pi a^3 \epsilon/3\, (=0.58\ \text{eV)}$.

Next,
we consider the activation  energies for vacancies in a face.
In this case, 
the energies of adjacent local minima 
are also supposed to be the same for simplicity.
Since the deformed regions are half of the inner case,
the deformation energies of the vacancies in the faces
are estimated to be
$\Delta E_{\text{face}\leftarrow\text{face} }
=4\pi a^3 \epsilon/3/2=\Delta E_{\text{inner}\leftarrow\text{inner}}/2
\, (=0.29\, \text{eV})$.

Moreover,
when a vacancy inside a face moves to an adjacent edge,
the deformation energy reduces to half of
$\Delta E_{\text{face}\leftarrow\text{face} }$,
since the deformed region is halved from that of
the transition in a face.
Hence,
we have
$\Delta E_{\text{edge}\leftarrow\text{face} }\approx
\Delta E_{\text{inner}\leftarrow\text{inner}}/4\ (=0.15\ \text{eV})$.
We see that
the estimated values  agree quantitatively  with the exact values.
Lastly,
when the vacancy in a face moves to an adjacent vertex,
the deformation energy is evaluated to be halved
to $\Delta E_{\text{edge}\leftarrow\text{face} }$.
Thus,
we have the following approximation: 
$\Delta E_{\text{vertex}\leftarrow\text{face} }\approx
\Delta E_{\text{inner}\leftarrow\text{inner}}/8(=0.07\ \text{eV})$,
which agrees qualitatively with
the exact value of $0.027$ eV.

Here,
we have revealed that
the activation energies for the system of $n_L=6$
are determined by the local deformation energies of $\Delta E^{\mathrm{loc.}}_{i,j}$ around the defects.
Accordingly,
the activation energies
are supposed to be almost independent of   the system size of $n_L$.
In fact,
we have
$\Delta E_{\text{inner}\leftarrow\text{inner}} =\Delta E_{(0,1,1)\leftarrow(0,0,0)}=0.58$ eV,
$\Delta E_{\text{face}\leftarrow\text{face}}=\Delta E_{(4,1,1)\leftarrow(4,0,0)}=0.33$ eV,
$\Delta E_{\text{edge}\leftarrow\text{face}}=\Delta E_{(4,4,0)\leftarrow(4,3,1)}=0.16$ eV,
and
$\Delta E_{\text{vertex}\leftarrow\text{face}}=\Delta E_{(4,4,4)\leftarrow(4,3,3)}=0.03$ eV
for $n_L=4$.
These results show that
all types of activation energies
are indeed almost independent of the system size
when $n_L \geqslant 4$.

On the other hand,
for $n_L=2$,
we have
$\Delta E_{\text{inner}\leftarrow\text{inner}}=\Delta E_{(0,1,1)\leftarrow(0,0,0)}=0.47$ eV, 
$\Delta E_{\text{face}\leftarrow\text{face}}=\Delta E_{(2,1,1)\leftarrow(2,0,0)}=0.39$ eV,
$\Delta E_{\text{edge}\leftarrow\text{face}}=\Delta E_{(2,2,0)\leftarrow(2,1,1)}=0.3$ eV,
and
$\Delta E_{\text{vertex}\leftarrow\text{face}}=\Delta E_{(2,2,2)\leftarrow(2,1,1)}=0.05$ eV,
which shows that
the uniform local deformation assumption for the activation energies
employed above
does not hold for $n_L=2$.
In other words,
the cluster of $n_L=2$ 
 is too small to separate the deformations of the surface 
from those of the central portion,
and thus
some non-negligible couplings are generated
between
the inner and surface deformations.
As a result of the couplings,
the relatively high
activation energies between inner transitions
are decreased,
while
the other relatively low activation energies
between surface transitions
are increased for $n_L=2$.

In addition,
the saddle connectivity graphs of
$n_L=4$, $6$, and $8$ depicted in Ref.~\cite{niiyama}
also show visually
that
the activation energies of
$\Delta E_{\text{inner}\leftarrow\text{inner}}
$,
$\Delta E_{\text{face}\leftarrow\text{face}}
$,
$\Delta E_{\text{edge}\leftarrow\text{face}}
$, and
$\Delta E_{\text{vertex}\leftarrow\text{face}}
$
are almost  independent of the sizes $n_L$ of the clusters.

\section{The second slowest relaxations\label{sec:2ndRelaxation}}

\begin{figure}[b]
\begin{center}
 \includegraphics[height=6.5cm]{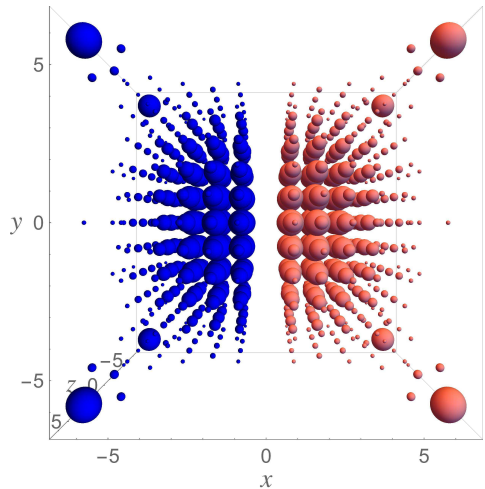}
\end{center}
\caption{%
We plot the second slowest relaxation mode of  $\vb*{P}_2$
in the same manner as in Fig.~\ref{fig:density}.
$\vb*{P}_2$ has the four-fold rotational symmetry around the $x$-axis.
The relaxation mode is polarized in the $x$-direction,
where
the excess  and shortage of probability
are distributed, respectively,
in $x>0$ and $x<0$, 
symmetrically with respect to the plane
of $x=0$.
Similarly to $\vb*{P}_2$,
$\vb*{P}_3$ and $\vb*{P}_4$
are the relaxation modes,
which are
polarized in 
the $y$- and  $z$-directions, respectively.
}
\label{fig:densityPx}
\end{figure}
In this section,
we examine the second slowest relaxations of 
$\lambda_2,\lambda_3,\lambda_4=-3.89\times 10^{5}\ \mathrm{s}^{-1}$.

In Fig.~\ref{fig:densityPx},
we plot  $\vb*{P}_2$ in the same way as in 
Fig.~\ref{fig:density}.
The probability deviations of 
$\vb*{P}_2$
are polarized in the $x$-direction.
Here,
the probability excess is in the region of $x>0$,
the probability shortage is in $x<0$,
and 
the probabilities are zero in $x=0$.
From this observation,
the relaxation process is expected as follows:
the probability excess moves in the opposite $x$-direction,
the probability shortage moves in the $x$-direction,
and 
these  pairs meet with each other in the region of $x=0$,
to be annihilated. 

To confirm this expectation, 
we evaluated the mean first passage times 
with sinks connected to $(0,n_y,n_z)$ for $-n_L\leqslant n_y,n_z \leqslant n_L$.
The longest passage time is
$t_{(2,0,0)}=4.05\times 10^{-6}\ \mathrm{s}$.
The resulting rate of this process
is
$2.47\times 10^{5}\ \mathrm{s}^{-1}$.
Also here,
the estimated values of the rate 
agree qualitatively with $|\lambda_2|$,
but they are somewhat smaller than
the exact rate of $|\lambda_2|=3.89\times 10^{5}\ \mathrm{s}^{-1}$,
because this approximate rate is evaluated
only from the longest mean first passage time,
as discussed in Sec.~\ref{sec:mfpt2vertex}.

Finally, we show that the approximate relation of 
$\lambda_2\approx 2 \lambda_1$ holds.
Here,
the value of $\lambda_2$  evaluated from 
the mean first passage time from $(2,0,0)$ to $x=0$
is approximated by that from $(2,0,0)$ to $(0,0,0)$.
From Fig.~\ref{fig:density}(b),
we see that
$\lambda_1$ is approximately evaluated from the mean first passage time
from $(0,0,0)$ to $(4,0,0)$,
which
is approximately twice as long as
that from  $(0,0,0)$ to $(2,0,0)$
since 
each of the transitions  requires almost the same transition time
as shown in Fig.~\ref{fig:FijFj}.
Hence,
the proportional relation
$1/|\lambda_1|:1/|\lambda_2|=4:2$ holds,
and thus
$\lambda_2=2\lambda_1$ holds.
Similarly,
we can derive
the approximate relations of 
$\lambda_3=2\lambda_1$ and  $\lambda_4=2\lambda_1$.

In this section,
we have confirmed that
the second slowest relaxations of
$\vb*{P}_2$,
$\vb*{P}_3$,
and 
$\vb*{P}_4$, respectively,
smooth out
the nonequilibrium distribution deviations
in the
$x$-,
$y$-,
and 
$z$-directions in the course of time,
and
the bottleneck processes for the second slowest relaxations 
are also
the slow diffusions inside the cluster.
This fact allows us to
derive the approximate relation of
$\lambda_2,\ \allowbreak\lambda_3,\ \lambda_4\approx 2 \lambda_1$.

\section{Symmetric evaluation of relaxation rates\label{sec:symmetric}}
\begin{figure*}[t]
\begin{center}
\includegraphics[height=5cm]{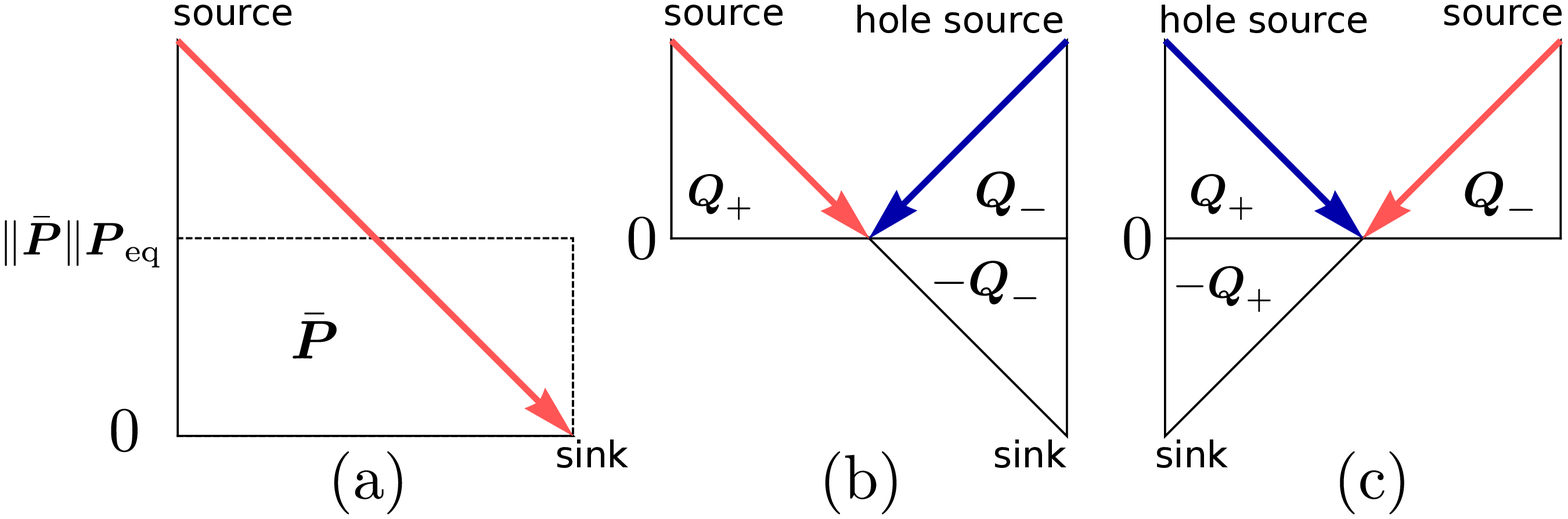} 
\end{center}
\caption{%
Schematic illustration of the  population methods.
(a) The stationary particle population of $(\bar{\vb*{P}})_i$
satisfying Eqs.~\eqref{eq:population}
and \eqref{eq:population2} is shown as a function of states $i$.
The red (light gray) arrow indicates how
the particles injected at the source move diffusively to the sink.
We also show $\|\bar{\vb*{P}}\| \vb*{P}_{\mathrm{eq}}
$  with a dashed line,
to illustrate the latter term in
Eq.~\eqref{eq:Qsiki0}.
(b)
The symmetric stationary population $\vb*{Q}$
with
a source and a sink connected to
the left and right end states, respectively,
is given by
Eq.~\eqref{eq:Qsiki0}.
The negative population $-\vb*{Q}_-$ is interpreted
as
the stationary population $\vb*{Q}_-$  of holes.
The blue (gray) arrow indicates how
the holes injected at the hole source move diffusively.
Particles from the source and
the holes 
meet at the center
and
are annihilated by pair annihilation.
(c)
The stationary population under the exchange of the sink and  source
is shown.
In this case,
particles of population $\vb*{Q}_-$
move to the left
and
holes of population $\vb*{Q}_+$
move to the right.
The particles and
the holes 
meet at the center
to be annihilated.
Both of the  mean annihilation times in (b) and (c)
agree with
the  mean first passage time, 
$\| \vb*{Q}_+ \|=\| \vb*{Q}_- \|$, 
of particles and holes.
(See the text.)
}
\label{fig:particle-hole}
\end{figure*}

\subsection{Symmetric population method \label{sec:symmetricpopulation}}
In the previous sections, 
we have successfully evaluated
the values of
the slowest and second slowest relaxation rates
with the use of the mean first passage times.
Recall that
the usages of the mean first passage times
for the slowest and second slowest relaxations
were different.
That is,
for the slowest relaxation,
the mean first passage times concerning  particles
were used,
whereas,
for the second slowest relaxations, 
the mean annihilation times of  particle-hole pairs 
were evaluated with the uses of the mean first passage times.

The difference manifests
itself in the symmetry of relaxation times under exchanging the sinks and sources.
Namely,
the approaches to evaluating
the slowest and second slowest relaxation rates,
respectively,
give
the relaxation times
that
are asymmetric and symmetric for exchanging
sinks and sources.
In fact, as shown in Eq.~\eqref{eq:P1relaxtime},
the slowest relaxation time is $7.33\times 10^{-6}$ s
with the source and the sinks being connected to the origin
and the vertices, respectively,
whereas,
with the sources and the sink being connected to the vertices and the origin,
the mean first passage time is given by
$\tau'=3.82\times 10^{-3}$ s
and thus
the relaxation time is $(\vb*{P}_\text{eq})_{(0,0,0)} \times \tau'=8.24\times 10^{-3}$ s.
The symmetry
corresponds to
the property
that
$\vb*{P}_i$ and $-\vb*{P}_i$
have the same value of $\lambda_i$,
and
hence it is required for a consistent treatment.

Here,
we  develop
an alternative population method for estimating mean first passage times
that is symmetric under the exchange of the sinks  and sources.

To this end,
we consider
the following stationary population equation:
\begin{align}
 K {\vb*{Q}}+\vb*{S}&= \vb*{0}, \label{eq:particle-hole}\\
\text{with}\  \| \vb*{Q}_+ \|= \|\vb*{Q}_-\|,&\quad
 \| \vb*{S}_+ \|= \|\vb*{S}_-\|=1. \label{eq:particle-hole2}
\end{align}
Here,
$
\vb*{Q}=
\vb*{Q}_+ - \vb*{Q}_-$
and
$
\vb*{S}=
\vb*{S}_+ - \vb*{S}_-,
$
where the positive population $\vb*{Q}_+$
and 
the negative population $-\vb*{Q}_-$
are given by
\begin{align}
\vb*{Q}_+=\frac{\vb*{Q}+|\vb*{Q}|}2,\quad
\vb*{Q}_-=-\frac{\vb*{Q}-|\vb*{Q}|}2,
\end{align}
with
$|\vb*{Q}|\equiv (|q_1|,|q_2|,\dots,|q_n| )$.
The source part $\vb*{S}_+$
and 
the sink part $-\vb*{S}_-$
are, respectively,
given by
\begin{align}
\vb*{S}_+=\frac{\vb*{S}+|\vb*{S}|}2,\quad
\vb*{S}_-=-\frac{\vb*{S}-|\vb*{S}|}2,
\end{align}
with
$|\vb*{S}|\equiv (|s_1|,|s_2|,\dots,|s_n| )$.
With use of
the stationary solution $\bar{\vb*{P}}$,
satisfying  Eqs.~\eqref{eq:population} and \eqref{eq:population2},
the stationary solution ${\vb*{Q}}$ of
Eqs.~\eqref{eq:particle-hole} and \eqref{eq:particle-hole2}
is
given by
\begin{align}
\vb*{Q}
=\bar{\vb*{P}}
- \| \bar{\vb*{P}} \|
\vb*{P}_{\mathrm{eq}},
\label{eq:Qsiki0}
\end{align}
In fact,
$\vb*{Q}$ given in Eq.~\eqref{eq:Qsiki0}
satisfies
the constraint
$\|\vb*{Q}_+\|=\|\vb*{Q}_-\|
$,
because
$
\|\vb*{Q}_+\|-\|\vb*{Q}_-\|=
\|\vb*{Q}\|
=\|
(\bar{\vb*{P}}-  \| \bar{\vb*{P}} \|
\vb*{P}_{\mathrm{eq}})
\|=\| \bar{\vb*{P}} \|-\| \bar{\vb*{P}} \|=0
$ holds.

As illustrated in Fig.~\ref{fig:particle-hole}(b),
the negative population $-\vb*{Q}_-$
can be interpreted as
the hole population of $\vb*{Q}_-$.
Hence,
the mean annihilation times of particles and holes
are, respectively,
given by
the first passage times of
$\|\vb*{Q}_+\|$
and
$\|\vb*{Q}_-\|$
as discussed in Sec.~\ref{sec:2ndRelaxation}.
Similarly,
$\vb*{S}_-$ is interpreted as
a hole source part that adds one hole per unit time.
Hence,
the constraint of $\|\vb*{S}_+\|=\|\vb*{S}_-\|=1$
means 
that
$\vb*{S}_+$
adds one particle  per unit time,
and $\vb*{S}_-$
adds  one hole  per unit time.
Note that,
since
$\vb*{P}_\text{eq}$ satisfies the detailed balance condition,
the particle flows of $\vb*{\bar{P}}$ and $\vb*{Q}$ are the same,
and so are their dominant pathways,
as illustrated in Figs.~\ref{fig:particle-hole}(a)
and \ref{fig:particle-hole}(b).

The constraint $\|\vb*{Q}_+ \|=\|\vb*{Q}_-\| $  means that
the mean first passage times are
symmetric under exchanging the sinks and sources.
In fact,
by exchanging the sinks and sources,
$\vb*{S}$ and 
$\vb*{Q}$
are, respectively,
converted to $-\vb*{S}$
and
$-\vb*{Q}$,
and  hence
$\vb*{Q}_+ $ and $\vb*{Q}_-$
are,
respectively,
converted to
$\vb*{Q}_- $ and $\vb*{Q}_+$,
as illustrated in Fig.~\ref{fig:particle-hole}(c).
Hence,
the mean first passage times
of Eqs.~\eqref{eq:particle-hole} and \eqref{eq:particle-hole2}
satisfy
$\|\vb*{Q}_- \|=\|\vb*{Q}_+\| $ with sinks and sources exchanged,
which 
is the same value as the value before the exchange.

Here,
with this symmetric population method,
we evaluate the slowest relaxation rate of $\lambda_1$
for the vacancy diffusion model of the KCl nanocluster. 
Setting
\begin{align}
(\vb*{S}_+)_{(0,0,0)}=1,\ 
 (\vb*{S}_-)_{(\pm n_L,\pm n_L,\pm n_L)}=1/8,
  \label{eq:SinP1}
\end{align}
and otherwise $(\vb*{S}_{\pm})_{i,j,k}=0$,
we evaluated
the symmetric stationary solution of Eq.~\eqref{eq:Qsiki},
thereby
obtaining
the slowest relaxation time
of
$\|\vb*{Q}_+\|=\|\vb*{Q}_-\|=
7.92 \times 10^{-6}$~s.
Namely,
with this symmetric method,
we obtained an approximation of the exact slowest relaxation time
of $-1/\lambda_1=5.2\times 10^{-6}$ s,
which is
symmetric under exchanging the sinks and sources
and
as accurate 
as the asymmetric result of $7.33\times 10^{-6}$ s
given in Sec.~\ref{sec:1stRelaxation}.

For the second slowest relaxation,
we set
\begin{align}
 (\vb*{S}_+)_{(2,0,0)}=1,\
(\vb*{S}_-)_{(-2,0,0)}=-1,    
 \label{eq:SinP2}
\end{align}
 and otherwise $(\vb*{S}_{\pm})_{i,j,k}=0$,
thereby
obtaining the symmetric result of
$\|\vb*{Q}_+\|=\|\vb*{Q}_-\|=
	4.05 \times 10^{-6}$ s,
	which of course agrees with the result given in Sec.~\ref{sec:2ndRelaxation}.

In this subsection,
we have developed
the symmetric population method for
mean first passage times,
which enables us to
approximately
evaluate 
the slowest relaxation times
symmetrically by exchanging the sinks and sources.

  \subsection{Symmetric population method as an inverse power method
  \label{sec:iteration}}

Here,
we show that
the iterative use of  the symmetric population method
enables us to
compute the slowest relaxation times accurately.

First,
the $n\times n$ matrix 
$P=(\vb*{P}_\text{eq},\vb*{P}_1,\vb*{P}_2,\dots,\vb*{P}_{n-1})$
is invertible,
where
$\vb*{P}_i$ is the eigenvector corresponding
to the $i$th relaxation mode.
We expand $\vb*{S}$ and $\vb*{Q}$ as
\begin{align}
\vb*{S}&=s'_0 \vb*{P}_\text{eq}+s'_1 \vb*{P}_1+s'_2 \vb*{P}_2+\dots,\\
\vb*{Q}&=q'_0 \vb*{P}_\text{eq}+q'_1 \vb*{P}_1+q'_2 \vb*{P}_2+\dots,\label{eq:Qsiki}
\end{align}
where the coefficients $s'_i$ and  $q'_i$ are defined as follows:
\begin{align}
 s'_i=(P^{-1}\vb*{S})_i,\quad  q'_i=(P^{-1}\vb*{P})_i.
 \label{eq:s'q'}
\end{align}
With the use of $s'_i$ and  $q'_i$,
Eq.~\eqref{eq:particle-hole}
is represented as
 \begin{align}
q'_0&= s'_0=0,\label{eq:q0s0}\\  
q'_i &=\frac{s'_i}{-\lambda_i} \quad (i=1,2,\dots).\label{eq:qisi}
 \end{align}
Substituting
Eqs.~\eqref{eq:q0s0} and  \eqref{eq:qisi}
into Eq.~\eqref{eq:Qsiki},
we have
\begin{align}
\vb*{Q}=\frac{s'_1}{-\lambda_1} \vb*{P}_1+
\frac{s'_2}{-\lambda_2} \vb*{P}_2+\dots.
\label{eq:Qsiki2}
\end{align}
Equations \eqref{eq:s'q'} and \eqref{eq:Qsiki2}
define 
the procedure to obtain $\vb*{Q}$ from $\vb*{S}$,
which
is denoted as $\vb*{Q}=Q(\vb*{S})$.

Then,
the mean first passage time approximation $\tau$ of the relaxation time
is written as follows:
\begin{align}
\tau=\frac{\|\vb*{Q}_+\|}{\|\vb*{S}_+\|}
=\frac{\|\vb*{Q}_-\|}{\|\vb*{S}_-\|}
= \frac{\|\vb*{Q}\|}{\|\vb*{S}\|}
= \frac{\|Q(\vb*{S})\|}{\|\vb*{S}\|},
\end{align}
as discussed in the previous subsection.

\begin{figure}[htb]
\begin{center}
\includegraphics[width=6cm]{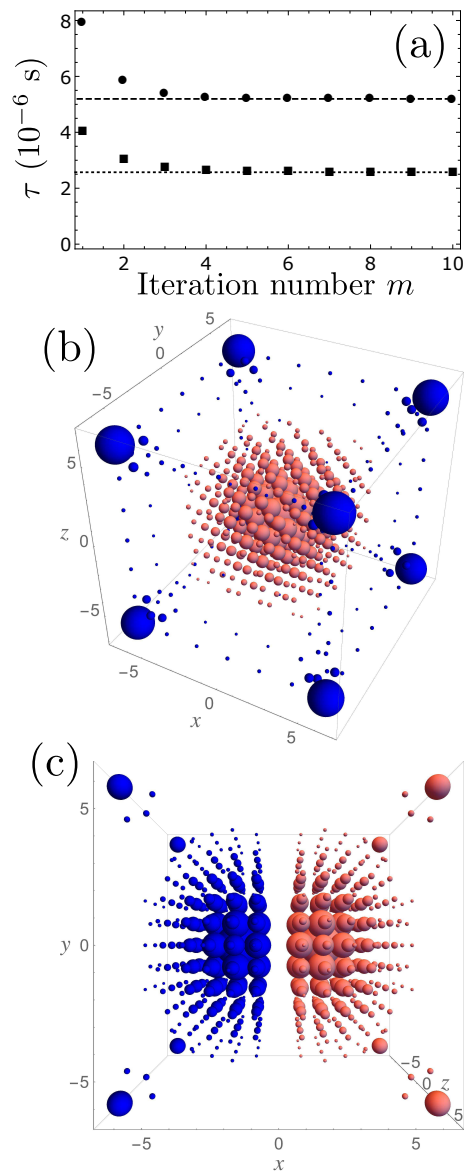} 
\end{center}
\caption{%
The mean first passage times of the symmetric population method
converge to  the slowest relaxation times.
(a)
The mean first passage times,
 $\tau =\|{Q}^{(m)}(\vb*{S})\|/\|{Q}^{(m-1)}(\vb*{S}) \|$,
 are plotted
 as a function of iteration number $m$ ($m=1,2,\dots,10$)
with circles
 for the slowest relaxation
and
with squares
for the second slowest relaxation.
The dashed and dotted lines show
 the values of the exact relaxation times of 
$1/|\lambda_1|$ and $1/|\lambda_2|$,
 respectively.
$Q(\vb*{S})$
are plotted in the  same manner as in Fig.~\ref{fig:density}(a),
 for the slowest relaxation  [panel (b)]
and 
for the second slowest relaxation [panel(c)].
}
\label{fig:iteration}
\end{figure}

Now,
we consider the effect of
the iterative use of this procedure.
We apply the $m$-times function composition of $Q$ to  $\vb*{S}$,
and
the resultant vector  ${Q}^{(m)}(\vb*{S})$ is given by 
\begin{align}
{Q}^{(m)}(\vb*{S})=\frac{s'_1}{(-\lambda_1)^m} \vb*{P}_1+
\frac{s'_2}{(-\lambda_2)^m} \vb*{P}_2+\dots,
\label{eq:Qsiki3}
\end{align}
which shows that, as $m\to \infty$,
if $s'_1\neq 0$, then $\vb*{Q}\propto \vb*{P}_1$, 
else if
$s'_1=\dots=s'_{i-1}= 0$ and $s'_i\neq 0$,
then $\vb*{Q}\propto \vb*{P}_i$. 
From this, we understand that
the symmetric population method
can be interpreted as an inverse power method for the eigenvalue problem \cite{matrixcomp}.

In Fig.~\ref{fig:iteration}(a), we plot the mean first passage times
\begin{align*}
\tau=
\frac{\|Q^{(m)}(\vb*{S})\|}{\|Q^{(m-1)}(\vb*{S})\|}
= 
\frac{\|Q (Q^{(m-1)}(\vb*{S}))\|}{\|Q^{(m-1)}(\vb*{S})\|}
\end{align*}
for  the slowest and second  slowest relaxation modes
with the settings  of Eqs.~\eqref{eq:SinP1} and \eqref{eq:SinP2},
respectively.
At $m=1$,
both are larger than the corresponding exact relaxation times,
because
$\vb*{S}$ 
of Eqs.~\eqref{eq:SinP1} and \eqref{eq:SinP2}
are selected so as to maximize
the mean first passage times of $\|Q(\vb*{S})\|/\|\vb*{S}\|$.
At $m=2$ and  $3$,
the mean first passage time approximations almost converge to
the corresponding relaxation times.
These findings
show that
the mean first passage time approximations of
the slowest and the second slowest relaxation times
satisfy
$s'_1\neq 0$ and 
$s'_1= 0, s'_2\neq 0$, respectively.
Moreover,
both of $\vb*S$ 
are sufficiently accurate,
so as to converge to the slowest and second slowest relaxation modes,
respectively,
with a few iterations.
Namely,
$\vb*S$ 
of Eqs.~\eqref{eq:SinP1} and \eqref{eq:SinP2}
closely approximate the exact eigenvectors of $\vb*{P}_i$ ($i=1,2$),
although $\vb*{S}$ are
the drastic simplifications of $\vb*{P}_i$
with very few sinks and sources.

To see the convergence of $Q^{(m)}(\vb*{S})$
to the eigenvectors 
with the symmetric population method,
we plot ${Q}(\vb*{S})$ for 
the slowest and  second slowest relaxation modes
in  Figs.~\ref{fig:iteration}(b) and \ref{fig:iteration}(c), respectively.
Comparing these graphs with
Figs.~\ref{fig:density}(a) and \ref{fig:densityPx},
respectively,
we see that
${Q}(\vb*{S})$ with Eqs.~\eqref{eq:SinP1} and \eqref{eq:SinP2}
are
almost the same
with
$\vb*{P}_1$
and 
$\vb*{P}_2$.
That is,
we can obtain accurate approximate eigenvectors
by applying this procedure just once to 
the quite simplified sinks and sources of $\vb*{S}$.

We
remark finally that
when
it is difficult to set $\vb*{S}$ to be in the convergence region of $\vb*{P}_2$,
we can obtain the first and  second slowest relaxation modes
simultaneously 
by iteratively applying $Q$ to two vectors that span a 
two-dimensional subspace and  orthogonalizing the vectors,
as in the general diagonalization algorithms
\cite{matrixcomp}.

\section{summary\label{sec:summary}}

We studied 
the  slowest and second slowest relaxations
of 
vacancy diffusion in a KCl nanocluster.

In Sec.~\ref{sec:1stRelaxation},
we found
that
the slowest relaxation mode of  $\vb*{P}_1$
has cubic symmetry around
the origin 
$(0,0,0)$,
where 
the excess probability at around the origin 
flows into the vertices of  $(\pm n_L,\pm n_L,\pm n_L)$
over the course of time evolution.
We also found 
that
the dominant pathways that carry large diffusive flows
are classified into two types of pathways from the origin to the vertices.
One is
through the face centers,
and 
the other
is through the edge centers.

To understand why these  pathways are selected as dominant pathways,
we estimated the mean first passage times from various states
to the vertex sinks.
As a result,
the surface diffusion turned out to be about $10^4$ times faster than
the surface diffusion  
at  room temperature of $k_\mathrm{B}T=0.03$ eV.
Hence,
the dominant pathways turned out to be
the shortest pathways to the surfaces.
There,
we  also gave an approximation of the slowest relaxation rate $\lambda_1$
with the use of the mean first passage times.

Next,
the reasons for the slow inner and
fast surface diffusions
were studied in terms of  the free energy landscape.
The sequences of free energies at  minima and saddle points
along the two types of pathways
were  examined. 
We found
that
the activation free energies
in  the inner region 
are
about twice as large as
those in  the surface region,
which
explains
the drastic slow inner diffusion. 
We also gave
an intuitive explanation for the ratio of the activation  energies
that are leading terms of the activation free energies,
with the  use of the individual atomic energies of $V_k$.

In Sec.~\ref{sec:2ndRelaxation},
we considered the second slowest relaxation modes.
With use of
the three-dimensional plot of Fig.~\ref{fig:densityPx},
the second relaxation modes of 
$\vb*{P}_2$,
$\vb*{P}_3$,
and 
$\vb*{P}_4$
turned  out to correspond
to relaxation of
the excesses and the deficiencies of probability
in the $x$-
$y$-, and 
$z$-directions, respectively.

The second slowest relaxation rate of $\lambda_2$
is also successfully estimated by use of 
the mean first passage times with sinks connected to the region of $x=0$.
There,
the intuitive explanation for
the approximate relation
$\lambda_2,\lambda_3,\lambda_4 \approx 2\lambda_1$
was given
in terms of  the free energy landscapes along the dominant pathways.

In Sec.~\ref{sec:symmetric},
we have developed a symmetric population method,
which computes the approximate relaxation rate
as the mean first passage times of particles and holes.
The symmetric population method
has a 
reasonable
property in that
both $\vb*{P}_i$ and $-\vb*{P}_i$ are the eigenvectors of
the same eigenvalue.
We have also shown that
iterative use of the symmetric population method
enables us to obtain the accurate slowest relaxation times,
similarly to the inverse power method of matrix diagonalization.

In summary, 
we have shown that 
the properties of the slowest relaxation modes are reconstructed 
by mean first passage times 
in Markov state models suitably connected with sinks and sources. 
The mean first passage times are useful 
to extract the bottleneck processes buried in Markov state models. 
We have also shown that 
the formation of the bottlenecks can be understood 
from the physical basis of potential energy landscapes 
that support the networks of the Markov state models.

\begin{acknowledgements}
The authors are very grateful to Shoji Tsuji and Kankikai
for the use of their facilities at Kawaraya during the early stage of
this study.
Y.S and
T.O. 
are supported by a Grant-in-Aid for Challenging Exploratory Research
(Grant No. JP 15K13539) from the Japan Society for the Promotion of Science.  
\end{acknowledgements}

\end{document}